\documentclass[reqno,11pt]{amsart}
\usepackage{amsmath,amsfonts,amsgen,amstext,amsbsy,amsopn,amsthm, amssymb}

\newtheorem{Lemma}{Lemma} \newtheorem{Theorem}{THEOREM}
 
 \theoremstyle{definition}

\newcommand\nn\nonumber

 \newcommand{\R}{\mathbb{R}}

\newcommand{\E}{\mathcal{E}}

\begin{document}

\title{The scattering length at positive temperature}

\author[B. Landon]{Benjamin Landon} 

\author[R. Seiringer]{Robert Seiringer} \address{Department of Mathematics and Statistics, McGill
  University, 805 Sherbrooke Street West, Montreal, QC H3A 2K6,
  Canada} 
\email{benjamin.landon@mail.mcgill.ca, robert.seiringer@mcgill.ca}

\date{November 7, 2011}

\begin{abstract}
  A positive temperature analogue of the scattering length of a
  potential $V$ can be defined via integrating the difference of the
  heat kernels of $-\Delta$ and $-\Delta + \frac 12 V$, with $\Delta$ the
  Laplacian. An upper bound on this quantity is a crucial input in the
  derivation of a bound on the critical temperature of a dilute
  Bose gas \cite{SU}. In \cite{SU} a bound was given in the case of
  finite range potentials and sufficiently low temperature. In this
  paper, we improve the bound and extend it to potentials
  of infinite range.
\end{abstract}

\thanks{\copyright\, 2011 by  
  the authors. This paper may be reproduced, in its entirety, for
  non-commercial purposes.}

\maketitle

\section{Introduction and Main Results}

Let $\Delta$ denote the usual Laplacian on $\R^d$, and let $V\geq 0$
be a multiplication operator on $L^2(\R^d)$.  An important ingredient
in the upper bound on the critical temperature for a dilute Bose gas
derived in \cite{SU} is a bound on the integral of the difference of
the heat kernels of $-\Delta$ and $-\Delta + \frac 12 V$. For
$\beta>0$, let
$$
g(\beta) = \frac 1 \beta \int_{\R^{2d}} \left( e^{2\beta \Delta} - e^{\beta(2\Delta - V)} \right)(x,y) \, dx\, dy \,,
$$
which is well-defined since the integrand is non-negative, by the
Feynman-Kac formula. It was shown in \cite[Lemma~V.1]{SU} that
$g(\beta)$ is equal to
$$
 \inf_{\phi \in H^1(\R^d)} \left\{ \int_{\R^d} \left( 2 |\nabla \phi(x)|^2 + V(x) |1-\phi(x)|^2 \right) dx + \frac 1 \beta \left\langle \phi \left| f\big( \beta(-2\Delta + V) \big) \right| \phi \right\rangle \right\} \,,
$$
where $f(t) = t(1-e^{-t})/(t-1+e^{-t})$. This variational principle
was used in \cite[Lemma~V.2]{SU} to derive an upper bound on
$g(\beta)$ for finite range potentials $V$ and $\beta$ sufficiently large. 
The function $f$ satisfies $1\leq f(t)\leq 2$ for all $t\geq
0$. In particular, one can replace $f$ by $2$ for an upper bound.

The functional under consideration is thus
\begin{equation}
\E_\beta(\phi) = \int_{\R^d} \left( 2 |\nabla \phi(x)|^2 + V(x) |1-\phi(x)|^2 + \frac 2 \beta |\phi(x)|^2 \right) dx\,.
\end{equation}
We assume that $V$ is radial and that $V\geq 0$. We are interested in
\begin{equation}
e(\beta) = \inf\left\{ \E_\beta(\phi) \, : \, \phi\in H^1(\R^d)\right\}\,.
\end{equation}
We shall assume that $V$ has finite {\em scattering length}
$0< a < \infty$ (whose definition will be recalled in the next
section). No regularity or integrability assumptions have to be imposed,
however. In particular, $V$ is allowed to have a hard core, i.e., we
allow $V(x)$ to be $\infty$ for $|x|\leq r$ for some $r\geq 0$. The
potential $V$ could also be a measure, e.g., a sum of
$\delta$-functions.

Our main result is the following.

\begin{Theorem}\label{thm1}
For $d=3$, 
\begin{equation}\label{3db}
e(\beta) \leq 8\pi a \left( 1 + \frac a{\sqrt{3\beta}} \right)^2\,.
\end{equation}
For $d=2$, 
\begin{equation}\label{2db}
e(\beta) \leq \frac {8\pi}{\ln\left( 1 + \beta/a^2\right)} \left( 1+ \frac{1+ a^2/\beta}{2 \ln\left( 1 + \beta/a^2\right)}\right)\,.
\end{equation}
\end{Theorem}

Analogous bounds can be derived for $d=1$ and $d>3$. Since the bounds
have applications in physics \cite{SU} only when $d=2$ or $d=3$, we
shall restrict our attention to these cases for simplicity. The proof
of Theorem~\ref{thm1} will be given in Sections~\ref{sec:proof3} and
\ref{sec:proof2} below.

If one is interested in bounds involving only the scattering length of
$V$, the bounds of Theorem~\ref{thm1} are optimal in a certain
sense. This will be further discussed in Section~\ref{sec:hc} where we
evaluate $e(\beta)$ in the case of a hard core potential.

\section{Scattering Length}

As in \cite{LY,LSSY}, the scattering length $a_R$ of the finite range
potential $V \chi_{\{|x|\leq R\}}$ is defined via the minimization problem
\begin{equation}\label{def:lr}
\lambda(R) = \inf \left\{  \E_\infty(\phi)  \, : \, \phi(x) = 0 \ \text{for $|x|>R$} \right\}\,.
\end{equation}
For $d=3$, we have, by definition,
\begin{equation}\label{d3l}
\lambda(R) = \frac {8\pi a_R}{ 1 - a_R/R} 
\end{equation}
while for $d=2$
\begin{equation}\label{d2l}
\lambda(R) = \frac {4\pi}{ \ln (R/a_R)}\,. 
\end{equation}
It is important to note that $a_R$ is independent of $R$ in case $V$
has finite range less than $R$. Note also that $0\leq a_R\leq R$ and that
$a_R$ is increasing in $R$. The scattering length of $V$ is then
defined to be $a = \lim_{R\to \infty} a_R$. The following simple
criterion for finiteness holds.

\begin{Lemma}\label{lem:fin}
The scattering length $a=\lim_{R\to \infty} a_R$ is finite if and only if
\begin{equation}
\int_{|x|>b} V(x) dx < \infty \qquad \text{($d=3$)}
\end{equation}
\begin{equation}
\int_{|x|>b} V(x) \left[ \ln (|x|/b)\right]^2 dx < \infty \qquad \text{($d=2$)}
\end{equation}
for some $b>0$. 
\end{Lemma}

The proof of this lemma will be given in Section~\ref{sec:ainf}.

\section{Proof of Theorem~\ref{thm1} in Three Dimensions}\label{sec:proof3}
It was shown in \cite{LY} that there is a unique minimizer $\psi_R$
for (\ref{def:lr}). The function $\psi_R$ is monotone decreasing,
radial, and satisfies 
\begin{equation}
2 \Delta \psi_R(|x|) = V(x) ( 1- \psi_R(|x|)) \quad \text{for $|x|\leq R$}
\end{equation}
in the sense of distributions (where the right side is interpreted as $0$ if $\psi_R = 1$ and $V=\infty$).  Moreover,
for $d=3$ the bound
\begin{equation}\label{bdp}
1\geq 1 - \psi_R(|x|) \geq  \max\left\{ \frac{ 1 - a_R/|x|}{1-a_R/R} \, , \,0\right\} \quad \text{for $|x|\leq R$}
\end{equation} 
holds. We also have  
\begin{equation}
\int_{|x|\leq R} V(x) ( 1 - \psi_R(|x|)) dx = 2 \int_{|x|\leq R} \Delta \psi_R(|x|) dx  = \frac{8\pi a_R}{1- a_R/R}\,.
\end{equation} 
From this identity and the monotonicity of $\psi_R$, we have the bound
\begin{align}\nonumber
\int_{R\leq |x|\leq R_1} V(x) dx  & \leq \frac 1{1-\psi_{R_1}(R)} \int_{R\leq |x|\leq R_1} V(x) (1-\psi_{R_1}(|x|) dx \\  & = \frac 1{1-\psi_{R_1}(R)} \frac{8\pi a_{R_1}}{1- a_{R_1}/R_1} -   \frac{8\pi a_{R}}{1- a_{R}/R} \label{3.3}
\end{align}
for $R_1> R > 0$. In the last step, we used the fact that $1-\psi_{R_1}$ and $1-\psi_{R}$ are proportional for $|x|\leq R$, and that $\psi_{R}(R) = 0$. Using, in addition, the bound (\ref{bdp}) and taking the limit $R_1\to \infty$, we obtain
\begin{equation}\label{bdv}
\int_{|x|\geq R} V(x) dx \leq \frac{8\pi a}{1-a/R}  - \frac{8\pi a_{R}}{1 - a_{R}/R}
\end{equation}
for $R>a$.

As a trial state for $\E_\beta$, we use the function $\psi_R$ for some $R>a$. Using (\ref{bdp}), we have
\begin{equation}
\int_{\R^3} |\psi_R(x)|^2 dx \leq \frac{4\pi a_R^3}{3} + \frac {a_R^2}{(1-a_R/R)^2} \int_{a_R\leq |x| \leq R} \left( 1/R - 1/|x|\right)^2 dx = \frac {4\pi a_R^2 R}{3} \,.
\end{equation} 
With the aid of (\ref{bdv}) and (\ref{d3l}) we hence obtain
\begin{equation}
\E_\beta(\psi_R) = \frac {8\pi a_R}{ 1 - a_R/R}  + \int_{|x|\geq R} V(x) dx + \frac 2 \beta \int_{\R^3} |\psi_R(x)|^2 dx \leq  \frac{8\pi a}{1-a/R} + \frac {8\pi a_R^2 R}{3\beta}\,.
\end{equation}
The choice $R = a+ \sqrt{3\beta}$, together with the bound $a_R\leq a$, yields our final result (\ref{3db}).

\section{Proof of Theorem~\ref{thm1} in Two Dimensions}\label{sec:proof2}
The proof for $d=2$ is similar to the three-dimensional case. Again the minimizer $\psi_R$ for (\ref{def:lr}) is monotone decreasing and radial, but now it satisfies
\begin{equation}\label{bdp2}
1\geq 1 - \psi_R(|x|) \geq  \max\left\{ \frac{\ln(|x|/a_R)}{\ln(R/a_R)} \, , \,0\right\} \quad \text{for $|x|\leq R$.}
\end{equation} 
Moreover, 
\begin{equation}
\int_{|x|\leq R} V(x) ( 1 - \psi_R(|x|)) dx = 2 \int_{|x|\leq R} \Delta \psi_R(|x|) dx  = \frac{4\pi}{\ln(R/a_R)}\,.
\end{equation} 
From this identity and the monotonicity of $\psi_R$, we thus have the bound
\begin{align}\nonumber
\int_{R\leq |x|\leq R_1} V(x) dx  & \leq \frac 1{1-\psi_{R_1}(R)} \int_{R\leq |x|\leq R_1} V(x) (1-\psi_{R_1}(|x|) dx \\  & = \frac 1{1-\psi_{R_1}(R)} \frac{4\pi }{\ln(R_1/a_{R_1})} -   \frac{4\pi }{\ln(R/a_{R})}
\end{align}
for $R_1> R > 0$. Inserting (\ref{bdp2}) and sending $R_1\to \infty$ yields
\begin{equation}\label{bdv2}
\int_{|x|\geq R} V(x) dx \leq \frac{4\pi}{\ln(R/a)}  - \frac{4\pi}{\ln(R/a_{R})}
\end{equation}
for $R>a$.

Again we use $\psi_R$ as a trial state for $\E_\beta$. 
From (\ref{bdp2}) it follows that
\begin{equation}
\int_{\R^3} |\psi_R(x)|^2 dx \leq  \frac {1}{[\ln(R/a_R)]^2} \int_{|x| \leq R} \left[ \ln(R/x) \right]^2 dx = \frac {\pi R^2}{2 [ \ln(R/a_R)]^2} \,.
\end{equation} 
With the aid of (\ref{bdv2}) and (\ref{d2l}) we hence obtain
\begin{equation}
\E_\beta(\psi_R) \leq  \frac{4\pi}{\ln(R/a)} +  \frac {\pi R^2}{\beta [ \ln(R/a_R)]^2}\,.
\end{equation}
If we choose $R=\sqrt{\beta}$ we thus obtain
\begin{equation}
e(\beta) \leq  \frac{8\pi}{\ln(\beta/a^2)}\left( 1 +  \frac {1}{2  \ln(\beta/a^2)}\right)
\end{equation}
for $\beta>a^2$.  To obtain a bound that holds for all $\beta$ we can
choose $R = a \sqrt{1+\beta/a^2}$ instead; this yields (\ref{2db}).

\section{The Hard Core Case}\label{sec:hc}
As an example, consider the case of a hard sphere potential of range $a>0$, i.e., $V(x)=\infty$ for $|x|\leq a$ and $0$ otherwise. In this case, the minimizer of $\E_\beta$ is, for $d=3$, given by 
\begin{equation}
\psi(|x|)= \min\left\{ \frac a {|x|}  e^{-(|x|-a)/\sqrt{\beta}} \, , \, 1\right\}
\end{equation}
and hence
\begin{equation}
e(\beta) =  -  8\pi a^2 \psi'(a) +\frac{8\pi a^3}{3\beta} = 8\pi a\left(1+ \frac a{\sqrt{\beta}} + \frac{a^2}{3\beta}\right)\,. 
\end{equation}
This shows that, except for the value of the constant in the error term, our bound (\ref{3db}) is optimal for large $\beta$. To leading order, $e(\beta)$ equals $8\pi a$, and the relative error is bounded by $O(a/\sqrt{\beta})$.

For $d=2$, the minimizer of $\E_\beta$ for the hard sphere potential is
\begin{equation}
\psi(|x|)= \min\left\{ \frac{ K_0(|x|/\sqrt{\beta})}{K_0(a/\sqrt{\beta})} \, , \, 1\right\}\,,
\end{equation}
where $K_0$ is the modified Bessel function of $2^{\rm nd}$ kind. Hence
\begin{equation}\label{eb2dhc}
e(\beta) = -\frac{4\pi a}{\sqrt{\beta}} \frac{K_0'(a/\sqrt{\beta})}{K_0(a/\sqrt{\beta})} + \frac{2\pi a^2}{\beta}
\end{equation}
in this case. The function $t\mapsto - t K_0'(t)/K_0(t)$ behaves like
$(\ln(2/t)-\gamma+o(1))^{-1}$ as $t\to 0$, where $\gamma$ denotes
Euler's constant \cite[Eq.~9.6.13]{AS}. Again, our bound (\ref{2db}) reproduces the leading
order exactly, and gives the same order of magnitude for the error
term as (\ref{eb2dhc}).

\section{Finiteness of the Scattering Length}\label{sec:ainf}

In this section we shall prove Lemma~\ref{lem:fin}. 
Consider first the case $d=3$. On the one hand, it follows from (\ref{3.3})--(\ref{bdv}) that if $a<\infty$ then $\int_{|x|\geq b} V(x) dx < \infty$ for all $b>a$. On the other hand, 
if $\int_{|x|\geq b} V(x) dx < \infty$, then 
\begin{equation}
\frac{8\pi a_R}{1-a_R/R} \leq \frac{8\pi b}{1-b/R} + \int_{|x|\geq b} V(x) dx 
\end{equation}
for all $R>b$, as can be seen by using the trial function 
\begin{equation}
\phi(x) = \left\{ \begin{array}{ll} 1 & \text{for $|x|\leq b$} \\ \frac{b/|x|-b/R}{1-b/R} & \text{for $b\leq |x|\leq R$} \\ 0 & \text{for $|x|\geq R$.} \end{array} \right.
\end{equation}
Hence $a\leq b + (8\pi)^{-1} \int_{|x|\geq b} V(x) dx$. 

For $d=2$, we can use the trial function 
\begin{equation}
\phi(x) = \left\{ \begin{array}{ll} 1 & \text{for $|x|\leq b$} \\ \frac{\ln(R/|x|)}{\ln(R/b)} & \text{for $b\leq |x|\leq R$} \\ 0 & \text{for $|x|\geq R$} \end{array} \right.
\end{equation}
for $R>b$. This gives
\begin{equation}
\frac{4\pi}{\ln(R/a_R)} \leq \frac{4\pi}{\ln(R/b)} + \frac 1{[\ln(R/b)]^2} \int_{b\leq|x|\leq R} V(x) [\ln(|x|/b)]^2 dx \,.
\end{equation}
We can rewrite this inequality as 
\begin{equation}
4\pi \ln(a_R/b) \leq \frac{\ln(R/a_R)}{\ln(R/b)}\int_{b\leq |x|\leq R} V(x) [\ln(|x|/b)]^2 dx\,.
\end{equation}
If $\int_{|x|\geq b} V(x) [\ln(|x|/b)]^2 dx$ is finite, this implies that $a_R$ is bounded independently of $R$. 
Taking $R\to \infty$ we  obtain
\begin{equation}\label{rhs}
4\pi \ln(a/b) \leq \int_{|x|\geq b} V(x) [\ln(|x|/b)]^2 dx\,.
\end{equation}

To show that the finiteness of $a$ implies integrability of the right side of (\ref{rhs}), we can use $\psi_R$ as a test function for $a_b$, evaluated on a ball of radius $R$, for $R>b>a$. Then,
\begin{equation}
\frac{4\pi}{\ln(R/a_b)} \leq \frac{4\pi}{\ln(R/a_R)} - \int_{b\leq |x|\leq R} V(x) \left(1-\psi_R(x)\right)^2 dx  \,.
\end{equation}
Using (\ref{bdp2}) this bound implies that
\begin{equation}
4\pi \ln(a_R/a_b) \geq  \frac {\ln(R/a_b)}{\ln(R/a_R)}\int_{b\leq |x|\leq R} V(x) \left[\ln(|x|/a_R)\right]^2  dx  \,.
\end{equation}
Letting $R\to \infty$ we obtain
\begin{equation}
4\pi \ln(a/a_b) \geq \int_{|x|\geq b} V(x) [\ln(|x|/a)]^2 dx\,. 
\end{equation}
This completes the proof.

\bigskip

\noindent {\it Acknowledgments.} Partial financial support by NSERC is gratefully acknowledged.

%%%%%%%%%%%%%%%%%%%%%%%%%%%%%%%%%%%%%%%%%%%%%%%%%%%%%%%%%%%%%%%%%%%%%%%%

\end{document}